\documentstyle[11pt,aaspp4]{article}

\input{psfig.sty}

\newcommand\axpa {1E~2259$+$586}

\newcommand\axpd {1RXS~J170849.0$-$400910}

\newcommand\axpf {XTE~J1810$-$197}
\newcommand\axpi {CXOU~J164710.2$-$455216}
\newcommand\axpis {CXO~J1647}

\newcommand\suz {{\it Suzaku}}

\newcommand\cxo {{\it Chandra}}
\newcommand\xmm {{\it XMM-Newton}}
\newcommand\swift {{\it Swift}}

\def\lesssim{\mathrel{\hbox{\rlap{\hbox{\lower4pt\hbox{$\sim$}}}\hbox{$<$}}}}
\def\gtrsim{\mathrel{\hbox{\rlap{\hbox{\lower4pt\hbox{$\sim$}}}\hbox{$>$}}}}

\begin{document}

\title{The 2006 Outburst of the Magnetar CXOU~J164710.2$-$455216}

\author{
Peter~M.~Woods\altaffilmark{1,2},
Victoria~M.~Kaspi\altaffilmark{3,4}, 
Fotis~P.~Gavriil\altaffilmark{5,6}, and
Carol~Airhart\altaffilmark{1}
}

\altaffiltext{1}{Dynetics, Inc., 1000 Explorer Blvd. 
Huntsville, AL 35806; 
Peter.Woods@dynetics.com}
\altaffiltext{2}{National Space Science and Technology Center, 320 Sparkman Dr. 
Huntsville, AL 35805}
\altaffiltext{3}{Department of Physics, Rutherford Physics Building, McGill 
University, 3600 University St., Montreal, QC, H3A~2T8, Canada}
\altaffiltext{4}{Astronomy Department, California Institute of Technology, 1200 E. California Blvd., Pasadena, CA 91125}
\altaffiltext{5}{NASA Goddard Space Flight Center, Astrophysics Science
Division, Code 662, Greenbelt, MD 20771}
\altaffiltext{6}{Center for Research and Exploration in Space Science and
Technology, University of Maryland Baltimore County, 1000 Hilltop Circle,
Baltimore, MD 21250}

\begin{abstract}

We report on data obtained with the {\it Chandra, XMM-Newton, Suzaku} and {\it
Swift} X-ray observatories, following the 2006 outburst of the Anomalous X-ray Pulsar
\axpi.  We find no evidence for the very large glitch and rapid exponential
decay as was reported previously for this source.  We set a $3\sigma$ upper
limit on any fractional frequency increase at the time of the outburst of
$\Delta\nu/\nu < 1.5 \times 10^{-5}$.  Our timing analysis, based on the
longest time baseline yet, yields a spin-down rate for the pulsar that implies
a surface dipolar magnetic field of $\sim 9 \times 10^{13}$~G, although this
could be biased high by possible recovery from an undetected glitch.  We also
present an analysis of the source flux and spectral evolution, and find no
evidence for long-term spectral relaxation post-outburst as was previously
reported.

\end{abstract}



\section{Introduction}

Of the various manifestations of isolated neutron stars, Anomalous X-ray
Pulsars (AXPs) and Soft Gamma Repeaters (SGRs) are the most dynamic members
within the class (see Woods \& Thompson 2006 and Mereghetti 2008 for
reviews).  The extraordinary changes they undergo in radiative output are
believed to be driven by the strong, evolving magnetic field which powers
their bright X-ray emission (Thompson \& Duncan 1995, 1996; Thompson,
Lyutikov \& Kulkarni 2002).  Now generally recognized as magnetars, AXPs and
SGRs are observed to have spin periods within a narrow range (2--12 s),
possess rapid spin-down rates indicative of their strong surface dipolar
field strengths (10$^{14}-10^{15}$ G), usually maintain X-ray luminosities of
10$^{33}-10^{36}$ ergs s$^{-1}$ though often with significant variability,
and at times emit bright, often super-Eddington bursts of X-rays and
gamm-rays.  These sudden bursts from magnetars are often clustered in events
referred to as ``outbursts.''

Although outbursts have been observed in SGRs since their discovery in 1979,
the first outburst from an AXP was not observed until 2002, when more than 80
individual bursts were recorded during a single 3 hour observation of \axpa\
(Kaspi et al.\ 2003).  These bursts were accompanied by a sudden spin-up
``glitch'' of fractional spin-frequency change $\sim 4 \times 10^{-6}$.  
Other examples of radiative outbursts accompanied by glitches have been seen
in AXPs 1E~1048.1$-$5937 (Dib, Gavriil \& Kaspi 2009) and 4U~0142+61
(Gavriil, Dib \& Kaspi 2009). Such outbursts are believed to be caused by a
fracturing of the neutron star crust, a result of internal magnetic stresses,
and include the external surface and magnetospheric disturbances that
follow.  The spin-up glitch and subsequent relaxation can be interpreted in
terms of angular momentum transfer from an initially more rapidly spinning
crustal superfluid to the crust, as mediated by unpinning and later
re-pinning of the superfluid angular momentum vortex lines.


The 2003 outburst of \axpf\ (Ibrahim et al.\ 2004) provided a second
opportunity to study an AXP outburst.  In this case, the AXP brightened by a
factor $\sim$300 directly post-outburst, as compared to the factor $\sim$20
flux increase observed for \axpa.   Not having been monitored prior to the
outburst,  it is not known if a spin-up glitch accompanied this event.


Several glitches in AXPs have been detected though with no obvious radiative
outburst.  These include glitches having comparable size to that seen in
\axpa\ at the time of its outburst (e.g. Kaspi \& Gavriil 2003, Dall'Osso et
al. 2003, Dib, Kaspi \& Gavriil 2008). A correlation between radiative
behavior and glitch activity has been claimed for \axpd\ (Rea et al. 2005,
Campana et al. 2007, Israel et al. 2007a) however Dib et al. (2008) argue
that the evidence for a correlation in this source thus far is marginal. Some
of the larger AXP glitches have recoveries that are unusual when compared
with those of rotation-powered pulsars, hinting at possible structural
differences with magnetars (Woods et al. 2004, Dib et al. 2008, Gavriil et
al. 2009, Livingstone et al. 2009).  Establishing the existence and
properties of AXP glitches, particularly for those for which there are
accompanying observable radiative changes, offers a view of the impact of the
internal event on the surface and immediate stellar surroundings, hence is
important.

In 2006, a third AXP outburst was detected, this time from \axpi.  This
outburst was signaled by a bright X-ray burst detected with the \swift\
observatory (Krimm et al.\ 2006) on 2006 September 21.  The AXP \axpi\
(\axpis\ hereafter) is a 10.6-s X-ray pulsar located in the young cluster of
massive stars, Westerlund 1 (Muno et al.\ 2006a).  Several
Target-of-Opportunity (ToO) observations with multiple X-ray telescopes were
performed following the burst detection.  Early \swift\ XRT (Campana \&
Israel 2006) and \xmm\ (Muno et al.\ 2006b) ToO observations showed that the
flux from the AXP increased by a factor $\sim$100 following the outburst --
similar in magnitude to the outburst from \axpf.  A fortuitous \xmm\
observation performed just 4 days prior to the \axpis\ outburst provided a
baseline for comparison.  Muno et al.\ (2007) showed that the spectrum of
\axpis\ hardened significantly when the flux increased and that the pulse
profile changed dramatically from a simple near-sinusoidal morphology to a
complex profile with three distinct peaks per cycle.  They argued that
currents in the magnetosphere of \axpis\ were excited during this event,
causing the enhanced radiative output.  These same currents could also
explain the change in pulse profile by altering the opacity of the
magnetosphere.  Muno et al.\ predict that as these currents relax and decay,
the pulse profile will return to its pre-outburst morphology.

From our first four \cxo\ ToO observations obtained in an interval 6--37 days
after the burst, we made a preliminary measurement of the  spin-down rate of
\axpis. (Woods et al.\ 2006).  From a separate phase-coherent timing analysis
of the \xmm\ and \swift\ data on \axpis, Israel et al.\ (2007b) claimed
evidence for an enormous glitch of magnitude $\Delta\nu/\nu \simeq 6 \times
10^{-5}$ at some point between the two \xmm\ observations bracketing the
outburst.  If correct, this glitch would represent the largest fractional
frequency change yet seen for any neutron star, hence would be of great
importance.  The frequency jump caused by the putative glitch was reported by
Israel et al. (2007b) to have decayed exponentially with an $e$-folding time of
$\sim$1.5 days, remarkable even by AXP standards.

Here, we report on a sequence of five \cxo\ ToO observations of \axpis\ from
six days following the outburst to 2007 February 02.  Also included in our
analysis are archival data from three \xmm\ observations, one observation
with \suz\, and 15 observations with the \swift\ XRT.  Combining these data,
we report on the flux decay, spectral evolution, pulse morphology changes,
and pulse timing of \axpis\ during the first several months of its
post-outburst recovery.  As we show in this paper, our analysis  reveals no
evidence for the previously claimed glitch in \axpis.

\section{X-ray Observations}

As part of our ongoing \cxo\ ToO program for AXPs we observed \axpis\ on five
separate occasions with the \cxo\ ACIS detector (Weisskopf et al.\ 2000)
between 2006 September 28 and 2007 February 02.  All observations utilized
the ACIS-S3 chip operated in Continuous Clocking (CC) mode.  This detector
mode provides only a one-dimensional image, but very fine time resolution
(2.85 ms) and thus a larger dynamic range of measureable source intensities. 
In Table~\ref{tab:obs}, we list relevant details of the \cxo\ observations
including the source exposure times, time of the observations, detector mode,
and observation database reference numbers.


\begin{table}[!h]
\begin{minipage}{1.0\textwidth}
\begin{center}
\caption{X-ray observation log for \axpi\ between 2006 September and 2007 
February. \label{tab:obs}} 
\vspace{10pt}
\begin{tabular}{ccccccc} \hline \hline

 Name & Observatory & Detector & Obsid & Time Relative to &  Date$^{a}$ & Exposure   \\
      &             &          &       & \swift\ Burst    &  (MJD TDB)  & (ksec)     \\
      &             &          &       & (days)           &             &            \\\hline
 
  Obs1 &     {\it XMM} &  EPN &   0404340101 & $-$4.01 & 53995.06 & 40.7 \\ 
  Obs2 &   {\it Swift} &  XRT &  00030806001 & 0.79 & 53999.85 & 1.9 \\ 
  Obs3 &   {\it Swift} &  XRT &  00030806002 & 1.55 & 54000.62 & 0.8 \\ 
  Obs4 &     {\it XMM} &  EPN &   0311792001 & 1.64 & 54000.70 & 26.8 \\ 
  Obs5 &   {\it Swift} &  XRT &  00030806003 & 2.01 & 54001.07 & 4.9 \\ 
  Obs6 &  {\it Suzaku} &  XIS &    901002010 & 2.62 & 54001.69 & 154.8 \\ 
  Obs7 & {\it Chandra} & ACIS &         6724 & 6.31 & 54005.38 & 15.2 \\ 
  Obs8 & {\it Chandra} & ACIS &         6725 & 11.04 & 54010.11 & 20.2 \\ 
  Obs9 &   {\it Swift} &  XRT &  00030806006 & 11.57 & 54010.64 & 2.0 \\ 
 Obs10 &   {\it Swift} &  XRT &  00030806007 & 12.49 & 54011.56 & 2.0 \\ 
 Obs11 &   {\it Swift} &  XRT &  00030806008 & 14.99 & 54014.05 & 2.2 \\ 
 Obs12 & {\it Chandra} & ACIS &         6726 & 18.36 & 54017.42 & 25.2 \\ 
 Obs13 &   {\it Swift} &  XRT &  00030806009 & 18.78 & 54017.85 & 3.5 \\ 
 Obs14 &   {\it Swift} &  XRT &  00030806010 & 18.99 & 54018.05 & 2.8 \\ 
 Obs15 &   {\it Swift} &  XRT &  00030806011 & 24.32 & 54023.38 & 5.6 \\ 
 Obs16 &   {\it Swift} &  XRT &  00030806012 & 30.17 & 54029.24 & 5.5 \\ 
 Obs17 &   {\it Swift} &  XRT &  00030806013 & 36.69 & 54035.76 & 2.8 \\ 
 Obs18 & {\it Chandra} & ACIS &         8455 & 37.32 & 54036.39 & 15.2 \\ 
 Obs19 &   {\it Swift} &  XRT &  00030806014 & 120.15 & 54119.21 & 2.1 \\ 
 Obs20 &   {\it Swift} &  XRT &  00030806015 & 123.06 & 54122.13 & 3.8 \\ 
 Obs21 & {\it Chandra} & ACIS &         8506 & 134.86 & 54133.92 & 20.2 \\ 
 Obs22 &     {\it XMM} &  EPN &   0410580601 & 149.41 & 54148.47 & 17.3 \\ 
 Obs23 &   {\it Swift} &  XRT &  00030806016 & 208.32 & 54207.38 & 4.3 \\ 
 Obs24 &   {\it Swift} &  XRT &  00030806017 & 209.16 & 54208.23 & 2.2 \\ 

\hline\hline
\end{tabular}
\end{center}
\begin{small}
\noindent$^{a}$ Mid-point of observation. \\
\end{small}
\end{minipage}\hfill
\end{table}


By coincidence, \axpi\ was observed just four days prior to the outburst on
2006 September 21 by \xmm\ (Muno et al.\ 2007).  We report on this observation
and on two additional observations obtained on 2006 September 22 and on 2007
February 17 (see Table 1).   In particular, we report on results from the PN
detector which was operated in Full Window (FW) mode for the first two
pointings and Large Window (LW) mode during the 2007 observation (Jansen et
al.\ 2001).  The PN-FW mode offers two-dimensional imaging and 73-ms time
resolution.  The LW mode also has two-dimensional imaging, but with finer time
resolution (48 ms) and thus a larger dynamic range.  The details of these
observations are contained in Table~\ref{tab:obs}. 

Following the annnouncement of burst activity from \axpis, \suz\ also
declared a ToO and observed the AXP within 2.5 days of the \swift\ burst
detection (Table~\ref{tab:obs}).  Here, we utilize data recorded by the four
XIS detectors onboard \suz\ (Koyama et al. 2007) operated in 1/8 window
mode.  The XIS detectors time tag photons to within 1.0~s and provide
two-dimensional spatial information.  The point-spread-functions of the four
\suz\ X-ray Telescopes (XRTs) matched to the XIS detectors have half-power
diameters $\sim$2$^{\prime}$.

Finally, the \swift\ XRT detector (Burrows et al. 2005) observed \axpis\ 15
times between 2006 September 17 and 2007 April 18 (Table~\ref{tab:obs}). 
During these observations, the detector operated in either Photon Counting
(PC) or Windowed Timing (WT) mode depending upon the source brightness and
pre-compiled commands to the spacecraft.  In WT mode, the XRT detector
records a one-dimensional image with fine time resolution (2.2 ms), analogous
to \cxo\ CC mode.  In PC mode, two-dimensional information is recorded, but
with very coarse time resolution (2.5 s) and a much more limited dynamic
intensity range.

\section{Temporal Analysis}
\label{sec:temporal}

For each of the five \cxo\ ACIS observations of \axpis, we started with the
standard level 2 filtered event list, using CIAO
v4.0\footnote{http://cxc.harvard.edu/ciao/}. First, we found the centroid for
the peak of the one-dimensional image from each CC-mode observation and
selected counts within 4 pixels of the centroid (i.e. $\pm$2$''$).  We
further selected counts with measured energies between 0.5 and 7.0 keV and
constructed light curves with 0.5 s resolution.  No bursts were observed in
any of the \cxo\ observations of \axpis.

Next, we converted the photon arrival times to the Solar system barycenter
using the CIAO tool {\tt axbary} and a source position of $\alpha$ = 16$^{\rm
h}$ 47$^{\rm m}$ 10$^{\rm s}$.2 and $\delta$ = $-$45$^{\circ}$ 52$^{\prime}$
16$^{\prime \prime}$.90 (J2000).  A simple FFT revealed a clear pulsed signal
at the spin frequency of \axpis.  The Fourier spectrum showed strong harmonic
content up to the third harmonic above the fundamental frequency.

Starting from the Observations Data Files, we constructed filtered event
lists for the two post-burst \xmm\ observations using the {\tt epchain} tool
provided within the \xmm\ Science Analysis Software (XMMSAS) v7.0.0 package. 
In the filtering process, we followed standard filtering procedures for PN
data and retained counts with patterns 0 through 12.  Next, we extracted
source and background event lists from circular regions of 35$''$ and 50$''$,
respectively.  The background regions were selected from the same CCD that
contained the source.  We further excluded time periods when a flare was
clearly visible in the background light curve and the average count rate in
the source region increased by more than 10\%.  Finally, the photon time tags
were corrected to the Solar system barycenter using the XMMSAS tool {\tt
barycen}.

For the one \suz\ observation, we utilized the four XIS cleaned event lists
produced by the standard pipeline analysis (v1.2.2.3).  Following the recipe
oulined in the \suz\ ABC
Guide\footnote{http://heasarc.gsfc.nasa.gov/docs/suzaku/analysis/abc/},  we
constructed a circular source region (1$'$ radius) centered on the \axpis\
position.  Note that a larger region that encompassed more of the source flux
could not be used because of the 1/8 window observing mode of the XIS
detectors for this observation.  The spatial filtering was performed using
the tool {\tt Xselect} and the four resulting source event lists were merged
into a single source event list using a custom Interactive Data Language
(IDL) procedure.  Next, we corrected a systematic
error\footnote{http:\//www.astro.isas.jaxa.jp\/suzaku\/analysis\/xis\/timing\/}
in the photon arrival times by adding 7.0~s to each time tag.  Finally, the
\suz\ HEASOFT tool {\tt aebarycen} was used to convert the photon arrival
times to the Solar system barycenter.

Fifteen \swift\ observations of \axpis\ were processed using HEASOFT v6.3.1. 
The Perl script {\tt cxrtpipeline} generated level 2 data for each data
set.   Source and background regions were defined for PC observations as
circles with radii of 47$''$ and 118$''$ respectively, with the background
region selected to be sufficiently distant from the source and other point
sources.  For WT observations, the source region was defined as a
40$\times$10 pixel  ($\sim$94$''\times$23$''$) rectangle, oriented along the
readout direction of the CCD.  Background regions of the same dimensions were
defined on either side of the source along the readout direction.  Source and
background event lists were extracted using {\tt Xselect} and the time tags
were then corrected to the Solar system barycenter using {\tt barycorr}.

\subsection{Pulse Ephemeris}

As discussed earlier, there is some disagreement regarding the spin evolution
of \axpis\ following the burst activity in 2006 September.  Our initial
report on a subset of the \cxo\ observations showed evidence for rapid
spindown (Woods et al.\ 2006), but not for exponential glitch recovery as was
claimed by Israel et al.\ (2006) based upon \xmm\ and \swift\ data.  A
subsequent publication by the same group (Israel et al. 2007b) included some
\cxo\ data and resulted in generally the same conclusion of a large amplitude
glitch ($\Delta\nu/\nu \simeq 6 \times 10^{-5}$) with a rapid exponential
glitch recovery on a time scale of $\sim$1 day.  Here, we consider all data
available from this time period in an effort to resolve this apparent
discrepancy.  Note that the discrepancy  applies only to the first $\sim$1
week post-burst due to the rapid decay of the exponential term in the Israel
et al.\ phase model.

In our analysis, we have invoked the commonly used technique of phase-coherent
timing (Woods et al.\ 2004).  The data were split into discrete segments
according to detector type and time such that there are no gaps in any one data
segment larger than one day.  Using these criteria, each \cxo, \xmm, and \suz\
observation was grouped into an individual data segment.  Some \swift\
pointings were combined with adjacent observations due to their close proximity
in time.  The last two \swift\ pointings were excluded due to their relatively
low signal-to-noise.  In total, 16 unique data segments were assembled.

Each data segment was folded using a pulse phase model defined as a Taylor
expansion of the phase $\phi(t)$ about a given epoch in time $t_0$ where
$\phi(t)=\phi(t_0)+\nu (t-t_0)+\frac{1}{2}\dot\nu (t-t_0)^2+...$.  The pulse
phase, frequency and frequency derivative at time $t_0$ are given by $\phi$,
$\nu$, and $\dot\nu$, respectively.  Initially, we set these model parameters
equal to the values determined by our earlier phase-coherent analysis of the
\cxo\ data (Woods et al. 2006).

\begin{figure}[!p]

\centerline{
\psfig{file=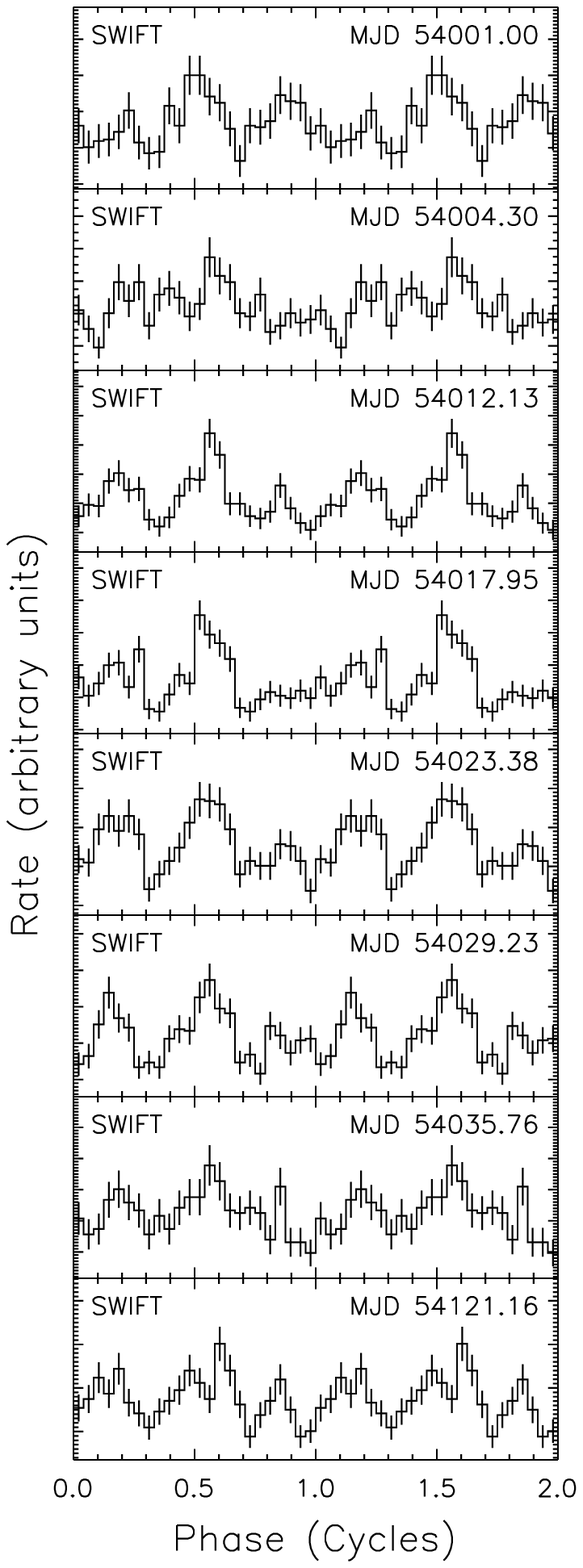,height=7.0in}
\psfig{file=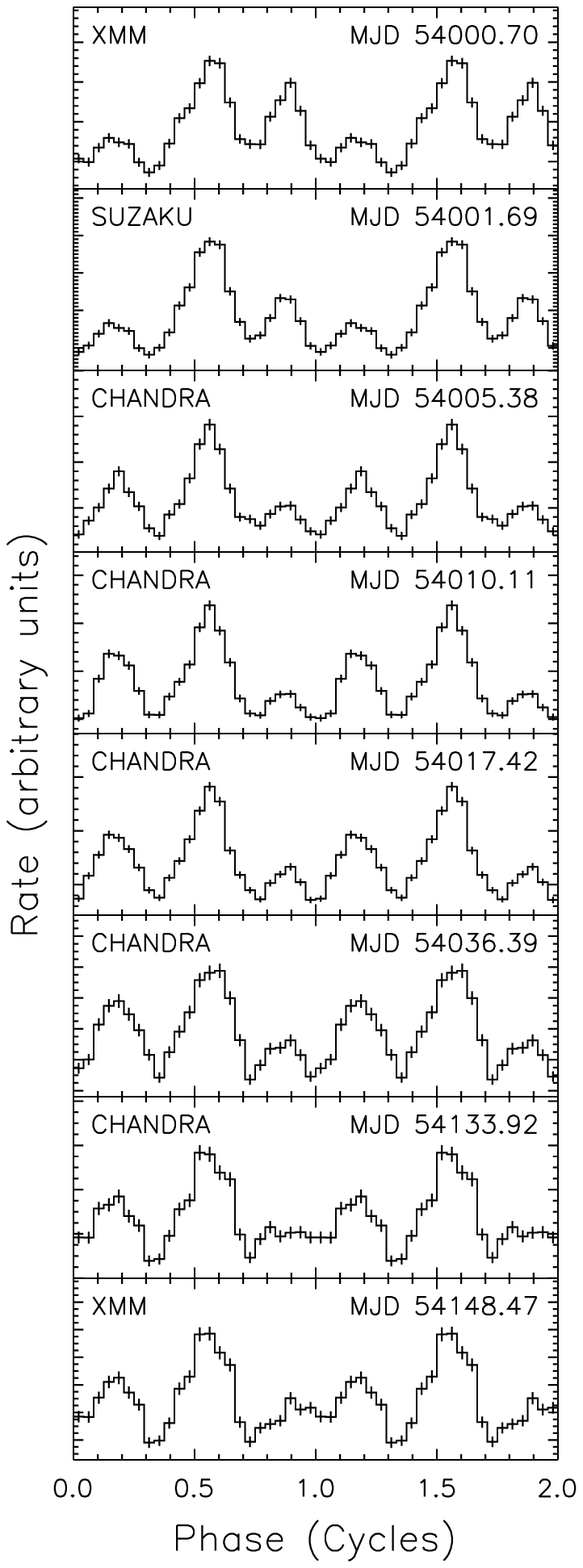,height=7.0in}
}

\caption{Left: Pulse profiles of \axpi\ between 0.5 and 7.0 keV from {\it
Swift} data.  Time progresses from top to bottom although not exactly at the
same rate in each column.  The total time durations are roughly the same.   The
profiles are phase aligned using our best-fit 3$^{\rm rd}$ order polynomial
ephemeris that includes no glitch. Note the difficulty in determining proper
phasing, particularly at early times, from the {\it Swift}-observed pulse
morphologies independently. Right:  Pulse profiles of \axpi\ between 0.5 and
7.0 keV during the post-burst time period for data from {\it Chandra, Suzaku,
and XMM}.  Time progresses from top to bottom.   The profiles are phase aligned
using our best-fit 3$^{\rm rd}$ order polynomial ephemeris that includes no
glitch. Due to the different instruments used for this comparison, count rate
units have been excluded.  Note the gradual change in pulse morphology from the
first pointing to the most recent.   \label{fig:profile}}

\end{figure}

When constructing the folded pulse profiles, we chose 32 phase bins per cycle
and selected counts within the energy range 0.5--7.0 keV to provide good
signal-to-noise and sufficient phase resolution for cross-correlation.  Each
of the 32 phase bins corresponds to $\sim$0.33 s in time.  The majority of
our data sets have time resolution much smaller than the size of our phase
bins, thus we are justified in assigning all counts in a given detector
accumulation interval to a single phase bin.  The \swift\ XRT data
accumulated in PC mode, on the other hand, have time resolution much coarser
than our phase bins.  For this reason, we have ``split'' these counts  across
multiple phase bins when folding the data.  This has the effect of smearing
out features in the pulse profile as we are effectively convolving a
rectangular function of phase width 0.33 cycles (the PC-mode time resolution
divided by the pulse period) with the pulse profile.  This smearing is
mitigated to some extent by the fact that we are averaging over several
hundred cycles at a time.  Nevertheless, the distortion of the pulse profiles
in the {\it Swift} data is not insignificant (see Fig.~\ref{fig:profile},
Left).

Folded pulse profiles for each data segment were then cross-correlated in the
Fourier domain with a high signal-to-noise pulse template derived from the
2006 \cxo\ observations.  The cross-correlation utilized amplitude and phase
information from the fundamental plus the first 3 harmonics of the Fourier
decomposition of the pulse profiles.

Visual inspection of the resulting initial phase residuals from an initial
simple quadratic model (i.e. fitting for $\phi(t_0)$, $\nu$, $\dot{\nu}$)
showed a number of outliers within the first few days of the burst activity,
but only, as expected, for the \swift\ XRT measurements.  All other data
points (in 2006) followed the quadratic trend.  In fact, the small number of
\swift\ outliers followed a similar quadratic trend, only this trend was
offset in phase from the other model by $\sim \frac{1}{3}$ cycles -- the
separation between the three peaks of the post-burst \axpis\ pulse profile.

The \swift\ pulse profiles (Fig.~\ref{fig:profile}, Left) have
signal-to-noise ratio significantly less than those of the other pulse
profiles (Fig.~\ref{fig:profile}, Right).   The pulsed signal was clearly
visible in these data, but in many cases the distinction between the three
peaks of the pulse profile was not clear.  Due to the generally very short
exposure times for the \swift\ XRT observations (see Table~\ref{tab:obs}),
the extensive usage of PC mode for these observations with coarse time
resolution, and the intrinsic variablity of the \axpis\ pulse profile, we
chose not to use these data to define our \axpis\ pulse ephemeris.

\begin{figure}[!p]

\centerline{
\psfig{file=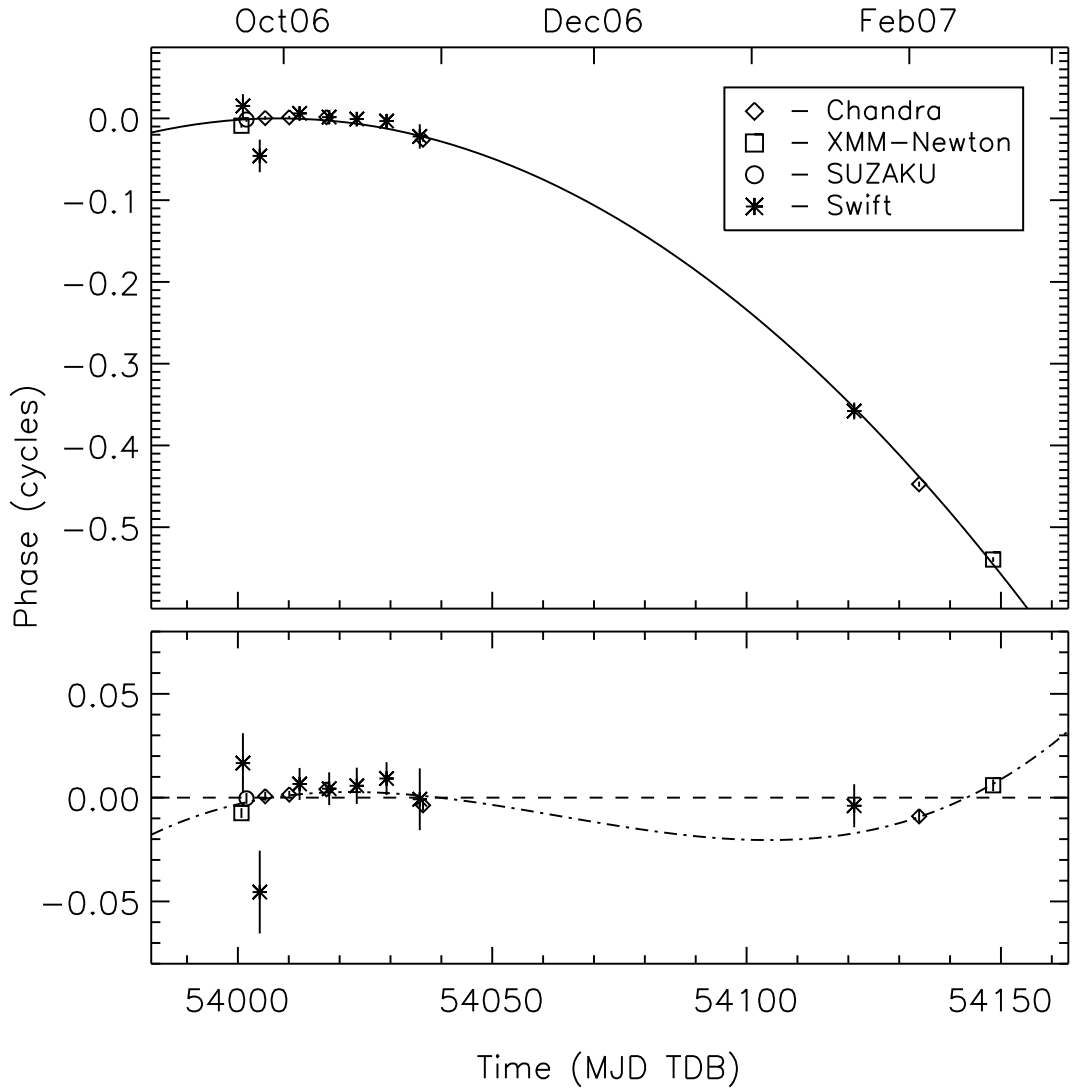,height=6.0in}}

\caption{Pulse phases of \axpi\ between 2006 September 23 and 2007 February
17.  {\it Top} -- Pulse phases minus a linear phase model.  The solid line
indicates a quadratic fit to these phase measurements.  {\it Bottom} -- Pulse
phases minus a quadratic trend.  The dash-dot line indicates a cubic fit to the
residuals. \label{fig:phases}}

\end{figure}

Next, we fit a quadratic phase model to only the \cxo/\xmm/\suz\ phases to
determine a new pulse ephemeris.  We repeated this procedure folding in the
2007 data, and generated a new pulse template using the \cxo\ and \xmm\
data.   The best-fit quadratic model is presented in Table~\ref{tab:ephem}
and shown graphically in Figure~\ref{fig:phases} (top). The high $\chi^2$
(31.4 for 6 degrees of freedom) is indicative of a residual systematic trend.
For this reason, we also tried including a cubic term in the phase fit and
the residuals became more acceptable ($\chi^2$ = 8.4 for 5 degrees of
freedom).  The resulting spin ephemeris for this cubic phase model is also
given in Table~\ref{tab:ephem} and Figure~\ref{fig:phases} (bottom).

For the quadratic fit, the measured $\dot{\nu} = -7.41(18) \times
10^{-15}$~Hz~s$^{-1}$, somewhat less than that measured by Woods et al.
(2006), $-1.38(28) \times 10^{-14}$ Hz~s$^{-1}$. It is also less than that
reported by Israel et al. (2007b). This is because of the longer baseline, and
the systematic trend toward a smaller spin-down rate at later times.  This
$\dot{\nu}$ implies a magnetic field,  calculated via $B=3.2 \times 10^{19}
(P\dot{P})^{1/2}$~G, of $9.5\times 10^{13}$~G.

For the cubic fit, however, the measured instantaneous spin-down rate is even
steeper, but is likely not stationary.  The average $\dot{\nu}$ over the
reported time span was $-6.5 \times 10^{-15}$ Hz s$^{-1}$. If correct, this
would imply a surface dipolar magnetic field of $B=8.9 \times 10^{13}$~G.

Thus, with the data reported here, we estimate the magnetic field strength at
$\sim 9 \times 10^{13}$~G, somewhat less than reported by Woods et al. (2006)
and Israel et al. (2007b).


\begin{table}[!h]
\begin{minipage}{1.0\textwidth}
\begin{center}
\caption{Spin Parameters for \axpi\ from 2006 September 23 through 2007 February
17. \label{tab:ephem}} 
\vspace{10pt}
\begin{tabular}{lc} \hline \hline
Start Observing Epoch (MJD) & 54000.692 \\
End Observing Epoch (MJD) & 54148.483 \\
Epoch (MJD TDB) & 54008.0000 \\\hline
Quadratic Fit: & \\\hline
Spin Frequency$^{a}$ , $\nu$ (Hz) & 0.0942448774(11) \\
Spin Frequency Derivative, $\dot{\nu}$ (Hz s$^{-1}$) & $-7.4(2) \times 10^{-15}$  \\\hline
Cubic Fit: & \\\hline
Spin Frequency, $\nu$ (Hz) & 0.0942448813(14) \\
Spin Frequency Derivative, $\dot{\nu}$ (Hz s$^{-1}$) & $-1.14(9) \times 10^{-14}$  \\
Second Spin Frequency Derivative, $\ddot{\nu}$ (Hz s$^{-2}$) & $8.5(18) \times 10^{-22}$\\ \hline\hline
\end{tabular}
\end{center}

\noindent$^{a}$ Numbers in parentheses represent 1$\sigma$ uncertainties
in the least significant digits quoted. \\
\label{ta:ephemeris}
\end{minipage}\hfill
\end{table}


Due to the extreme change in pulse profile from before the burst activity to
following it, we could not phase connect to the 2007 September 17 \xmm\
observation.  We measure a pulse frequency of 0.0942447(5) Hz during this
pre-burst observation, consistent with the values reported by Muno et al.\
(2006b) and Israel et al.\ (2007b).  An extrapolation of our post-burst spin
ephemeris (Table~\ref{ta:ephemeris}) to the time of the burst is 
consistent with the pre-burst spin frequency.  Thus, we place a 3$\sigma$ upper
limit on a glitch in \axpis\ at the time of the burst activity at
$\Delta\nu/\nu < 1.5 \times 10^{-5}$.  Our analysis of the complete sample of
X-ray data during the time surrounding the 2006 burst activity shows  no
evidence for a glitch or an exponential glitch recovery of the magnitude and
time scale as that reported by Israel et al.\ (2007b).

We believe that the discrepancy between our timing results and those reported by Israel et al.\
(2007b) amounts to cycle count ambiguities between observations immediately
following the burst activity and misidentification of peaks in the three-peaked
\axpis\ post-burst pulse profile for some of the \swift\ data sets.  Proving
the latter point would require a direct comparison between our pulse profile
correlation technique and the approach employed by Israel et al.   The former
issue can be more easily investigated by measuring frequencies during the
observations in close proximity to the burst activity.

As the frequency error ($\delta\nu$) in an individual observation is inversely
proportional to its duration,  $\delta\nu_j \propto T_j^{-1}$, the longer
observations are more suitable for comparison of our polynomial model to the
polynomial plus exponential model of Israel et al..  In addition, those
observations closest to the burst (i.e.\ ``glitch'') epoch are most
constraining due to the rapid decay of the putative exponential component.  For these
reasons, we made our frequency comparisons using the first \xmm\ (Obs4) and
\suz\ (Obs6) post-burst observations.

\begin{figure}[!h]

\centerline{
\psfig{file=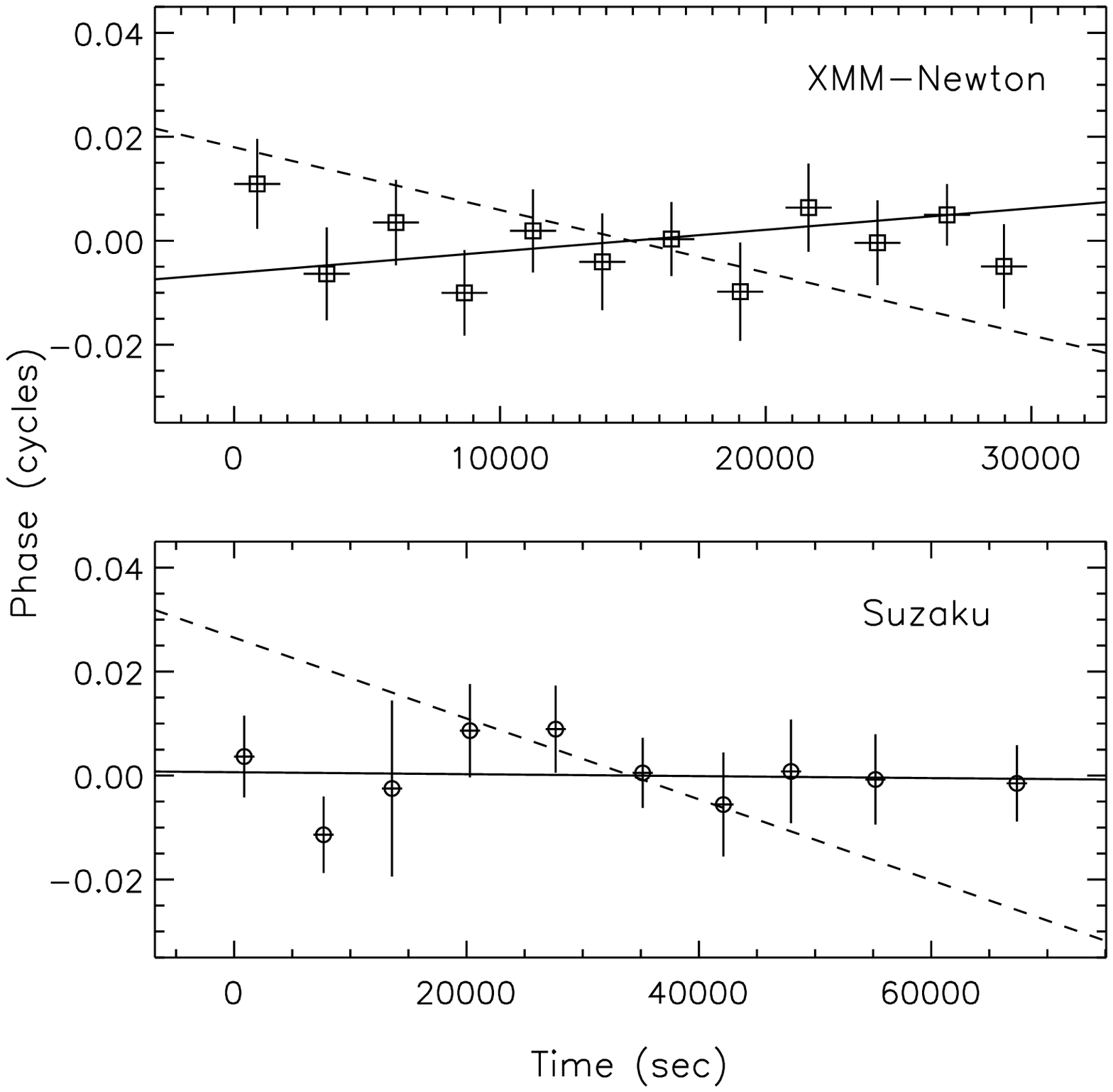,height=5.0in}}

\caption{Pulse phases of \axpi\ for the first \xmm\ (top panel) and \suz\
(bottom panel) observations following the burst detection on 2006 September
21.  The solid line indicates the expected local frequency for the 3$^{\rm rd}$
order polynomial model reported here (Table~\ref{tab:ephem}). The uncertainty
in the frequencies for the polynomial model (i.e.\ slope of the line) are
comparable to the thickness of the lines.  Note that there is no significant
difference in the instantaneous frequencies at these two epochs for the 3$^{\rm
rd}$ order 2$^{\rm nd}$ order polynomial models. The dashed line indicates the
polynomial plus exponential model of Israel et al.\ (2007b).  Frequency
uncertainties for this model cannot be inferred directly from the fit results
that have been reported.  \label{fig:freq}}

\end{figure}

Using the same phase-coherent timing approach as before, we divided the two
observations into 12 and 10 segments, respectively.  In order to limit the
effect of pulse profile changes, we constructed pulse templates from the
complete data set of each observation folded on the best fit local
frequency.  Phase offsets were measured for each segment and a linear phase
model was fit to the residuals to refine the pulse frequency.  We repeated
this procedure generating a new template profile and frequency measurement. 
The pulse phases for each data set are shown in Figure~\ref{fig:freq}.  Next,
we computed the pulse frequencies at each of these epochs for the two models
under consideration.  The difference in frequency between the model values
and the best-fit local frequency is shown graphically by the overplotted
lines in Figure~\ref{fig:freq}.  Our 3$^{\rm rd}$ order polynomial model is
represented by the solid line and the Israel et al.\ model is indicated by
the dashed line.


It is clear from Figure~\ref{fig:freq} that the frequencies predicted by our
model are more consistent with the measured data.  In particular, the
$\chi^2$/dof for the \xmm\ and \suz\ data sets evaluated using our phase
model are 9.4/10 and 5.1/8, respectively.  On the other hand, the Israel et
al.\ model returns much larger $\chi^2$/dof of 29.2/10 and 44.0/8,
respectively.  The probabilities that one would obtain $\chi^2$ values this
large by chance are 1 $\times$ 10$^{-3}$ and 6 $\times$ 10$^{-7}$,
respectively.  Thus, we conclude that the polynomial plus exponential glitch
model of Israel et al.\ (2007b) is inconsistent with the data.

\section{Spectral Analysis}

Using the source and background regions defined in Section \ref{sec:temporal},
X-ray pulse invariant spectra were constructed for each of the \cxo, \xmm\ and
\suz\ observations. Due to the low signal-to-noise ratio of the \swift\ data,
they were excluded from the spectral analysis. Response matrices were
constructed using {\tt mkacisrmf} from CIAO v4.1.2 and {\tt rmfgen} from XMMSAX
v7.1 for the \cxo\ and \xmm\ data, respectively. Pre-computed \suz\ response
matrices were utilized. The resulting energy spectra were grouped to ensure
that at least 25 source counts were contained within each energy bin.

\begin{figure}[!p]
\centerline{
\psfig{file=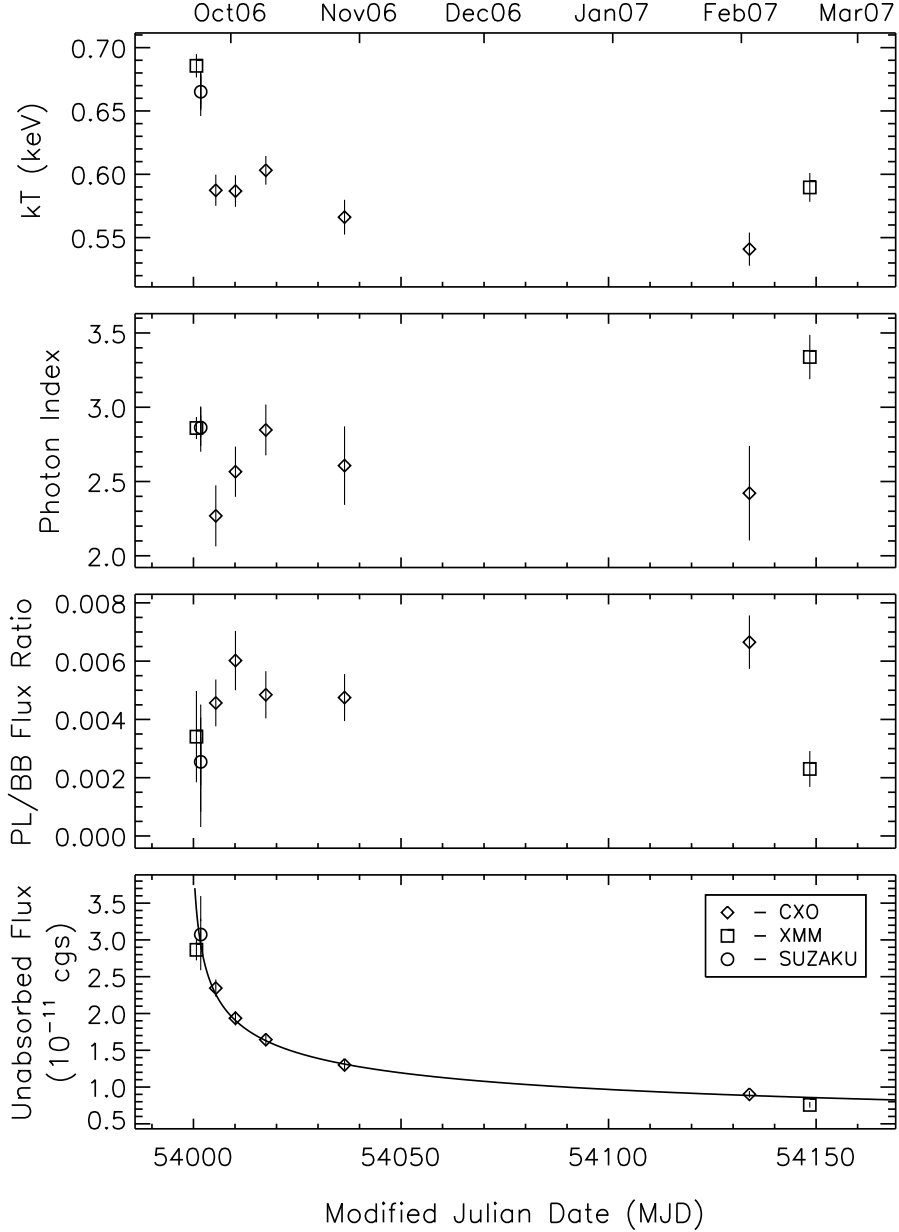,height=7.0in}}
\caption{Spectral history of \axpi\ between 2006 September 23 and 2007 February
17.  Shown from top to bottom are the blackbody temperature, photon index,
ratio of 2--10~keV power-law component flux to the bolometric blackbody flux,
and total unabsorbed 2$-$10 keV flux.  The power-law fit ($F = F_0 t^{-\alpha}$) to the
flux measurements is overplotted in the bottom panel.  The best-fit power-law
index was $-0.306 \pm 0.005$.  
\label{fig:spec}}
\end{figure}

The spectra were simultaneously fit to a model defined by the sum of a blackbody
and a power law modified by interstellar absorption. The spectra were fit using
XSPEC v11\footnote{http://heasarc.gsfc.nasa.gov/docs/xanadu/xspec/} where only
the interstellar absorption was forced to remain constant among data sets.  We
obtained a good fit to the data having a $\chi^2$ of 6208 for 6138 degrees of
freedom.  Fit parameter uncertainties were estimated using the {\tt error}
command in XSPEC.  Figure~\ref{fig:spec} shows the results of this fit,
specifically the temporal evolution of the blackbody temperature, photon index,
ratio of 2$-$10 keV power-law flux to bolometric blackbody flux, and the total
unabsorbed 2$-$10 keV flux. There is some indication of a slight drop in
blackbody temperature at times greater than $\sim$2 days following the outburst,
however, the spectral fit results show no significant long-term variability in
the spectral shape. The total unabsorbed 2$-$10 keV flux clearly decreases
rapidly following the outburst. The flux decay is well described by a power law
with a decay constant of $0.306 \pm 0.005$.




\section{Discussion}

Glitches in AXPs have now been shown to be generic to the class
(e.g. Kaspi, Lackey \& Chakrabarty 2000; Kaspi \& Gavriil 2003;
Dall'Osso et al. 2003), particularly at the times of large
radiative outbursts (Kaspi et al. 2003; Dib, Kaspi \& Gavriil 2008;
Gavriil, Dib \& Kaspi 2009).  Therefore, it would not be surprising
if \axpi\ suffered a glitch at the time of its 2006 outburst; indeed
given that every other well observed AXP radiative outburst
has included a timing anomaly, it might at first glance be surprising
that our analysis of the existing data reveals none, especially
given the previous claim by Israel et al. (2007b).

However it is important to note that the $3\sigma$ upper limit we place
on the amplitude of any glitch that occured is $\Delta\nu/\nu < 1.5\times10^{-5}$,
large by AXP (and indeed any) glitch standards.  Only one AXP glitch seen
thus far is higher (Dib et al. 2008).  Thus although our analysis rules
out the extremely large glitch claimed by Israel et al. (2007b), it does
not rule out a glitch having fractional amplitude similar to those seen
in most AXP glitches.  If \axpi\ had been subject to phase-coherent
timing prior to the event (impossible with e.g. the {\it Rossi X-ray Timing
Explorer} due to its pre-outburst faintness and a nearby, unrelated bright source), 
much smaller glitches would have been detectable.

The observed systematic deviation from simple spin-down, even in the
relatively small interval covered by our observations (see
\S~\ref{sec:temporal}), is interesting.  The magnitude of $\dot{\nu}$ appears
to have declined monotonically since the radiative outburst.  This is
suggestive of glitch recovery, even in the absence of direct evidence for a
glitch.   Strong glitch recoveries, with enhanced spin-down rates immediately
post-glitch (with or without an accompanying radiative event) have been
observed in other AXPs following large glitches (e.g. Kaspi et al.  2003;
Kaspi \& Gavriil 2003; Dall'Osso et al. 2003).  Their origin is unclear and
may be signalling structural differences between magnetars and lower-magnetic
field neutron stars (e.g. Dib, Kaspi \& Gavriil 2008).  On the other hand,
large variations in spin-down rate have been seen in other AXPs (e.g. Gavriil
\& Kaspi 2004; Dib, Gavriil \& Kaspi 2009) in the absence of glitches, so we
cannot conclude that glitch recovery is the origin of the $\dot{\nu}$
variation in \axpi.  

If it were glitch recovery, however, and $\dot{\nu}$ is recovering to a
smaller absolute value that is also its long-term average, then the true
$\dot{\nu}$ could be even smaller than is reported here. Indeed the
instantaneous $\dot{\nu}$ at the end of the time span reported here (using
the cubic phase model), if close to the long-term average, would imply a
surface dipolar magnetic field strength of $3.7 \times 10^{13}$~G,
surprisingly low. This would be lower than for any other known {\it bona
fide} magnetar (the previous lowest was for AXP 1E~2259+586 having $B=5.9
\times 10^{13}$~G) and below the so-called ``quantum critical field,'' $4.4
\times 10^{13}$~G. It is lower than for 8 known otherwise normal apparently
rotation-powered pulsars, including PSR J1847$-$0130 which has $B = 9.4
\times 10^{13}$~G (McLaughlin et al. 2003). Determining the long-term average
spin-down rate for \axpi\ requires regular monitoring over a much longer time
span during which the pulse profile remains stable.


Israel et al. (2007b) reported a decay in the flux of \axpi\ post-outburst
that is well modelled by a power law of index $-0.28 \pm 0.05$, consistent
with the index we observe, $-0.311 \pm 0.005$.  As they remark, this is
similar to what was observed for the flux decay of AXP 1E~2259+586 after its
2002 outburst (Kaspi et al. 2003; Woods et al. 2004; Zhu et al. 2008). 
However Israel et al. (2007b) further report that when spectrally decomposing
the emission, the power-law component  decayed more rapidly than the
blackbody component, with power-law indexes for these two components of
$-0.38\pm 0.11$ and $-0.14 \pm 0.10$, respectively. They argued that this
implied that the cooling time scale for hypothesized surface hot spots was
shorter than for relaxation of the region producing the power-law emission.
However, as shown in Figure~\ref{fig:spec}, when considering particularly the
{\it Chandra} data, we find that the ratio of power-law- to
blackbody-component flux remained relatively stable, or even increased
slightly after $\sim$140 days.  We do not understand the origin of this
observational discrepancy.  The two decaying in concert is consistent with
the picture of magnetospheric Compton scattering of enhanced surface thermal
emission (Thompson et al. 2002; Zane et al. 2009) in the presence of decaying
surface emission, with no change in the magnetospheric twist (e.g. \"Ozel \&
G\"uver 2007).

\section{Conclusions}

We have shown that there is no direct evidence to support the claim that AXP
\axpi\ in the massive star cluster Westerlund 1 exhibited a glitch at the
time of its 2006 radiative outburst, and we set a $3\sigma$ upper limit on
the magnitude of any such glitch of $\Delta\nu/\nu < 1.5 \times 10^{-5}$,
which is larger than those seen in most other AXP radiative outbursts.  We
show that a previous claim by Israel et al. (2007b), that a far larger glitch
occured, was a result of misidentification of correct pulse phases in low
signal-to-noise ratio data.  We further show that the spin-down rate of
\axpi\ is lower than was measured immediately post-event, suggesting possible
strong glitch recovery as has been seen in other AXPs, even in the absence of
direct evidence for a glitch.  The revised spin-down rate implies a surface
dipolar magnetic field strength of $\sim 9 \times 10^{13}$~G, although strong
glitch recovery could be biasing this result upwards.

\acknowledgments{\noindent {\it Acknowledgements} --  }  The authors wish to
thank the referee for useful comments and the CXC help desk for guidance on the
Chandra spectral analysis.  PMW is grateful for support from NASA through SAO
grant GO7-8077A. VMK thanks the California Institute of Technology for
hospitality  and acknowledges support from a Moore Scholarship, NSERC via a
Discovery Grant, CIFAR, FQRNT, the Canada Research Chairs Program, and the
Lorne Trottier Chair in Astrophysics and Cosmology.

\end{document}